\documentclass[accepted=2020-05-19, a4paper, aps,pra,twocolumn,superscriptaddress,longbibliography]{quantumarticle}

\pdfoutput = 1

\usepackage[utf8]{inputenc}

\usepackage{graphicx,bm,amsmath,eufrak,amssymb,url}
\usepackage[usenames,dvipsnames]{color}
\usepackage[bbgreekl]{mathbbol}
\usepackage{xcolor}
\usepackage{hyperref}

\usepackage[T1]{fontenc}
\usepackage[utf8]{inputenc}
\usepackage[english]{babel} 
\usepackage{verbatim}
\usepackage{graphicx}

\usepackage[numbers,sort&compress]{natbib}
\usepackage{bbold}

\newcommand{\be}{\begin{equation}}
\newcommand{\ee}{\end{equation}}
\newcommand{\bea}{\begin{eqnarray}}
\newcommand{\eea}{\end{eqnarray}}

\definecolor{smoothred}{HTML}{C5232F}
\definecolor{mygreen}{rgb}{0,0.5,0}
\definecolor{myblue}{rgb}{0,0,0.75}
\definecolor{mymagenta}{cmyk}{0,1,0,0.12}

\def\doi{http://dx.doi.org/}

\newcommand{\ict}{\affiliation{Center for Quantum Physics, Faculty of Mathematics, Computer Science and Physics, University of Innsbruck, A-6020, Innsbruck, Austria}}
\newcommand{\iqoqi}{\affiliation{Institute for Quantum Optics and Quantum Information of the Austrian Academy of Sciences, A-6020 Innsbruck, Austria}}

\makeatother

\begin{document}
\title{Variational quantum state preparation via quantum data buses}
\author{\href{mailto:viacheslav.kuzmin@uibk.ac.at}{Viacheslav V. Kuzmin}} \ict \iqoqi
\author{\href{mailto:pietro.silvi@uibk.ac.at}{Pietro Silvi}}\ict \iqoqi

\begin{abstract}
We propose a variational quantum algorithm to prepare ground states
of 1D lattice quantum Hamiltonians specifically tailored for programmable
quantum devices where interactions among qubits are mediated by Quantum
Data Buses (QDB). For trapped ions with the axial Center-Of-Mass (COM)
vibrational mode as single QDB, our scheme uses
resonant sideband optical pulses as resource operations, which are
potentially faster than off-resonant couplings and thus less prone
to decoherence. The disentangling of the QDB from the qubits by
the end of the state preparation comes as a byproduct of the variational
optimization. We numerically simulate the ground state preparation
for the Su-Schrieffer-Heeger model in ions and show that our strategy
is scalable while being tolerant to finite temperatures of the COM mode.
\end{abstract}
\maketitle

\section{Introduction}

Realizing many-body quantum states on quantum devices offers an experimental
pathway for studying the equilibrium properties of interacting lattice
models 
\cite{EndresAssemble, GrossCold, Dwave1800qubits}, quench dynamics~\cite{Hebenstreit2013,Jurcevic2014,Martinez2016,Pichler2016},
or it can be viewed as a quantum resource, e.g., for sensing~\cite{KitagawaSSS,SimoneSSS,RaphaelSSS,BalintSSS}.
Adiabatic state preparation has been experimentally demonstrated,
e.g.,~on trapped ions~\cite{MonroeAdiabatic} and atoms in optical
tweezers~\cite{Bernien2017}. However, this strategy is
currently limited by the finite coherence times of quantum devices,
incompatible with the long annealing times required by the adiabatic
criterion~\cite{Richerme2013}. 
An alternative, promising pathway towards quantum ground state preparation
is given by feedback-loop quantum algorithms, and especially by the
Variational Quantum Eigensolver~\cite{Peruzzo2014,Farhi2014,McClean2016,Kandala2017,VQSPontryagin2017,OptControl2011,Klco2018}
(VQE). In VQE,
parametrized $-$ generally non-universal $-$ quantum resource operations,
available on a given quantum device, are arranged in a variational
sequence, or circuit ansatz, to generate entangled trial states.
 The trial states approximate a target ground state as a result of
a closed-loop optimization of the variational parameters, where an
energy cost function is minimized. 

For trapped ion qubits, VQE was
recently demonstrated using as entangling resource either the M{\o}lmer-S{\o}rensen
gate \cite{BlattVQEIonMS2018}, or a programmable analog quantum simulator
of the long-range XY spin model \cite{Kokail2019}. Both these entangling
resources generate effective interactions between ion qubits by coupling
them to vibrational modes (phonons). Implementing the optical couplings
off-resonantly, one effectively achieves robustness of these resource
operations to the temperatures of the phonon modes. However, this
strategy yields relatively slow rates~\cite{Smith2016,Elben2019},
limiting the performance of VQE, as the quantum processing suffers from
decoherence.

\begin{figure}
\includegraphics[width=\linewidth]{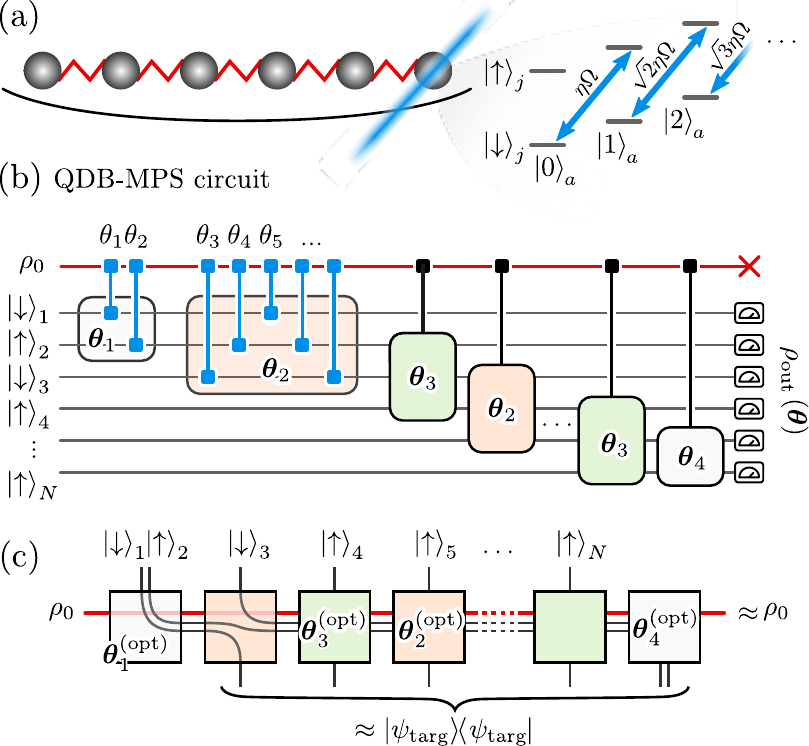}\caption{(a) Blue-detuned optical pulse acting on ion $j$ realizes
interaction $H_j$, Eq.~\eqref{eq:}, which couples the internal ion levels to levels of the COM phonon mode. The coupling rates depend on the state of the system. (b) QDB-MPS circuit
in an ion trap generating trial states as in Eq.~\eqref{eq:-56}. The COM mode (red line), as single QDB, is initially
cooled to a low-temperature state $\rho_{0}$, and ions (black lines)
are initialized in pure states. Variational operations $\text{exp}(-i\theta_{k}H_{j_k})$ (blue links)
are virtually grouped into boxes,  parametrized by corresponding vectors $\boldsymbol{\theta}_{i}=\{\theta_{k}\}$, as explicitly illustrated for the first two boxes. The out-coming
ions are measured, while the state of the QDB is discarded. (c) Mapping
of circuit (b) with the optimized parameters, $\{\boldsymbol{\theta}_{i}^{\text{(opt)}}\}$,
to an MPS diagram with the qubits state approximating $\left|\psi_{\text{targ}}\right\rangle $.}
\label{fig:-1}
\end{figure}

Here we propose a strategy to improve the efficiency
of VQE on programmable quantum devices where interaction between qubits
is mediated by auxiliary degrees of freedoms: Quantum Data Buses (QDBs).
This strategy employs interactions between qubits and QDBs, naturally available
in these devices, as variational resource operations in the quantum algorithm.
We illustrate our approach on trapped ions, using the acoustic Center-Of-Mass (COM) phonon mode
$-$ the axial vibrational mode where all ions oscillate in phase $-$ acting as a single QDB.
We employ resonant optical coupling of ion qubits with the COM mode,
yielding faster rates than off-resonant strategies,
and thus being less prone to decoherence.
Unlike logical gates designed with the resonant coupling, such as
the C-phase gate~\cite{Cirac1995}, our variational approach is
tolerant to finite temperatures of the COM mode, which do not limit the
state preparation fidelity.
Moreover, our strategy does not require to disentangle the QDB after
each resource operation, but only at the end of the state preparation,
ultimately yielding faster processing. The task of disentangling the
QDB from the qubits is carried out by the optimization algorithm itself:
As the output qubit state approaches the unique ground state of a
target non-degenerate Hamiltonian, the QDB becomes disentangled without
ever being measured.

\begin{figure}
\includegraphics[width=\linewidth]{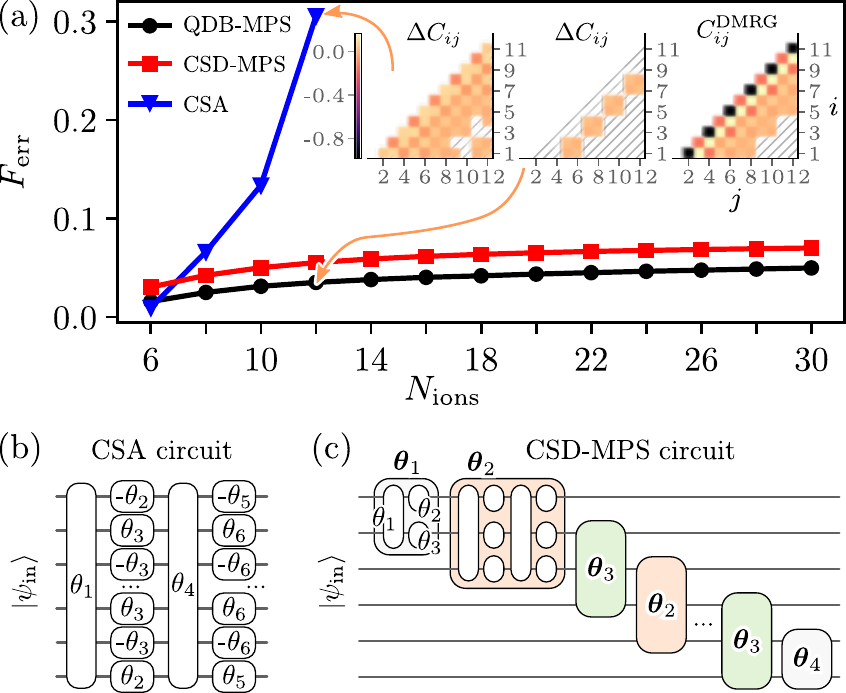}\caption{(a) Error function $F_{\text{err}}$~(\ref{eq:-3}) versus size $N_{\text{ions}}$
of the states prepared in the circuits QDB-MPS, CSA shown in (b),
and CSD-MPS shown in (c) for the fixed number of variational parameters
$N_{p}=18$. The insets show correlation matrix $C_{ij}^{\text{DMRG}}$~(\ref{eq:-1})
(right) with $i<j$ for $\left|\psi_{\text{targ}}\right\rangle $ with $N_{\text{ions}}=12$ obtained by the numerical
DMRG method, and deviations $\Delta C_{ij}=C_{ij}-C_{ij}^{\text{DMRG}}$
for $C_{ij}$ obtained for states prepared by the circuits CSA (left)
and QDB-MPS (middle); the arrows indicate the corresponding $F_{\text{err}}$
data-points. The dashed pixels represent arias with negligible absolute
value $<0.01$. (b) CSA and (c) CSD-MPS circuits generating trial states~\eqref{eq:-55}.
CSA circuit uses global all-to-all qubits operations~\eqref{eq:-6}
and singe-qubit rotations $\sigma_{i}^{z}$. CSD-MPS circuit uses
local entangling MS gates~\eqref{eq:-4} and singe-qubit rotations
$\sigma_{i}^{z}$.}
\label{fig:-9} 
\end{figure}

Via numerical simulations, we investigate the scaling and robustness
properties of our strategy for 1D lattice models. As a benchmark goal,
we aim to variationally prepare the ground state, $|\psi_{\text{targ}}\rangle$,
of the Su-Schrieffer-Heeger (SSH) Hamiltonian~\cite{SSHmodel1979,SSHmodel1983,AsbothSSH2016}
on ion qubits in a linear trap. Using only blue-detuned sideband optical
pulses~\cite{GardinerZollerBook} as resource operations, we design
the variational circuit ansatz shown in Fig.~\ref{fig:-1}, which
can efficiently realize Matrix Product States (MPS),
a class of tailored variational wavefunctions capable of accurately capturing
the equilibrium physics of many-body quantum systems in 1D \cite{MPS1,AreaLaw}.
This `QDB-MPS circuit' can incorporate various symmetries of the target
model for enhanced performance, including approximate translational
invariance in the bulk. 
In this paper, we will show that our approach is scalable, as we can
approximate 1D ground states at saturating precision in the system
size $N_{\text{ions}}$, without increasing the number of variational
parameters $N_{p}$ {[}see Fig.~\ref{fig:-9}(a){]}.
Additionally, we compare the results from the QDB-MPS circuit with
other VQE strategies, still designed for trapped ion hardware, but using different sets of (coherent) entangling resources:
{\it (i)} site-filtered M{\o}lmer-S{\o}rensen gates
{\it (ii)} an analog quantum simulator of a long-range XXZ model \cite{Kokail2019}.
We evaluate the respective accuracies in terms of various figures of merit, including excitation energy, fidelity,
and two-point correlations.
We demonstrate that
our strategy can realize highly-accurate ground states even for QDB
initialized at finite temperatures. Finally, we show that the QDB-MPS
circuit can be up-scaled beyond the single-trap limit by being implemented
in modular ion traps~\cite{Blatt2008,Monroe2013}. Overall, we
consider our strategy a viable, efficient route towards quantum state
preparation in ion traps as well as in other quantum platforms that
rely on QDBs, such as atom qubits coupled via photons in waveguides~\cite{Hood2016,Yu2019a}
or cavities~\cite{Kollar2017,Cohen2018,Vaidya2018,Welte2018},
or superconducting qubits coupled by microwave resonators~\cite{Fitzpatrick2017,Fink2017,Song2017,Schuetz2019qdb}.

The paper is organized as follows. In section \ref{sec:resources},
we describe the ion trap resource operations which we employ in the
QDB-MPS circuit. We also review the resources used by other VQE strategies in ion traps,
which we consider for comparison. We briefly introduce the target
SSH model in section \ref{sec:VQExplained}, and we discuss in detail the
VQE strategies. 
We exhibit the results of our numerical simulation in section \ref{sec:results}.
Finally, in section \ref{sec:outlook}, we summarize our conclusions
and argue future perspectives for our approach.

\section{Quantum resources in trapped ions}

\label{sec:resources}

We consider a setup of $N_{\text{ions}}$ ions in a linear Paul trap
as a relevant quantum platform. The system qubit, or spin 1/2, at
site $j$ is encoded by two internal electronic levels of ion $j$~\cite{GardinerZollerBook},
labeled $\left|\downarrow\right\rangle _{j}$ and $\left|\uparrow\right\rangle _{j}$.
Each ion, driven by laser fields in the transition between levels $\left|\downarrow\right\rangle _{j}\leftrightarrow\left|\uparrow\right\rangle _{j}$,
experiences a recoil associated with absorption or emission of photons.
This mechanism couples the internal ion levels with vibrational (phonon) modes,
either axial or radial, depending on the optical setup~\cite{GardinerZollerBook},
and allows phonons to be used as bosonic QDBs for the ion qubits.
In ion traps, as well as in other quantum hardware using QDBs~\cite{Shi2015,Hood2016,Yu2019a,Kollar2017,Cohen2018,Vaidya2018,Welte2018,Fitzpatrick2017,Fink2017,Song2017,Sillanpaa2007},
various strategies have been developed to construct entangling quantum
operations between qubits, while ensuring that the QDBs are disentangled from the
qubits by the end of each operation. In what follows, we review some
of these strategies.

\textit{Resonant Sideband Resources} $-$ One possible approach to
exploit QDBs in quantum hardware is to transfer information from one
qubit to another by populating QDBs with real excitations~\cite{Cirac1995,Sillanpaa2007,Shi2015,Song2017,Fink2017}.
In trapped-ion quantum computers, a typical choice is to elect the axial COM mode as single QDB, thanks to notable advantages:
The mode frequency is well separated from the other phonon modes thanks to the linear dispersion at $k=0$ of the acoustic band,
and additionally the qubit-phonon coupling is (roughly) homogeneous along the chain.
Using focused beams, each individual ion can be coupled to the COM mode, usually via sideband optical
pulses~\cite{Cirac1995} implemented in the Lamb-Dicke regime~\cite{GardinerZollerBook}
(where the Lamb-Dicke parameter $\eta$ is small, $\eta\ll1$). These pulses have frequencies
detuned from the atomic transition exactly by the frequency of the
COM mode $\nu$. For the case considered further in this paper, it is sufficient to
consider a single resonant sideband, either the $-\nu$ negative detuned (red sideband) or the $+\nu$ positive detuned (blue sideband).
As the two choices lead to equivalent results, here we only illustrate the blue sideband case for simplicity, which is described
by the Anti-Jaynes-Cummings Hamiltonian 
\begin{equation}
H_{j}=i\eta\Omega\left(a\sigma_{j}^{-}-a^{\dagger}\sigma_{j}^{+}\right)\label{eq:}
\end{equation}
addressing ion $j$, where $\Omega$ is the Rabi frequency,
and achievable rates are of the order $\eta\Omega\propto25\,\text{kHz}$~\cite{Lemmer2013,Debnath2018}.
In this picture, $a$ is the destruction operator for the COM mode, and $\sigma_{j}^{-}=(\sigma_{j}^{+})^{\dagger}=\left|\downarrow\right\rangle \!\left\langle \uparrow\right|_{j}$
is the lowering operator for ion $j$. In Eq.~(\ref{eq:}), we used
a gauge transformation $a\to e^{i\varphi}a$ to highlight the anti-unitary
symmetry $(\sigma^{\pm}\to\sigma^{\mp},a\to a^{\dagger},t\to-t)$
in the interaction Hamiltonian.

In Ref.~\cite{Cirac1995}, it was proposed to use resonant sideband
pulses to realize the C-phase gate in ion traps. However, the main hindrance
in using gates built out of resonant interactions with a bosonic QDB,
as Eq.~\eqref{eq:}, is the requirement for the QDB to be perfectly
initialized in a pure state, typically the vacuum $\left|0\right\rangle _{a}$,
where index $a$ refers to the QDB mode. This requirement stems from the
incommensurate rates $\sqrt{l+1}\eta\Omega$ of the transitions $\left|\downarrow\right\rangle _{j}\left|l\right\rangle _{a}\leftrightarrow\left|\uparrow\right\rangle _{j}\left|l+1\right\rangle _{a}$,
depending on the number of excitations $l$ stored in the QDB, as
highlighted in Fig.~\ref{fig:-1}(a). As a result, the gate fidelity
of resonant sideband gates in ion traps decreases roughly linearly
with temperature $T$ of the initial thermal state $\rho_{0}(T)=(1-e^{-1/T})\exp(-a^{\dagger}a/T)$
(in units of $k_{B}=\hbar=1$) of the COM mode.

\textit{Off-Resonant Sideband Resources} $-$ To address this hindrance,
alternative strategies to realize entangling operations were developed, in which couplings of the qubits to
QDBs are implemented off-resonantly, and the QDBs are only virtually populated~\cite{Molmer1999,PorrasCirac2004,Majer2007,Douglas2015,Vaidya2018}.
A prototypical example in ion traps, the M{\o}lmer-S{\o}rensen
(MS) gate~\cite{Molmer1999}, can be implemented by coupling a subset
$\mathcal{S}$ of ions to the COM mode by a bichromatic laser beam
with two frequencies, detuned from the qubit frequency by $\pm\left(\nu-\delta\right)$
with $\delta\ll\nu$. The resulting Hamiltonian
\begin{equation}
H_{\text{MS}}=\frac{(\eta\Omega)^{2}}{\delta}\sum_{i<j}^{\mathcal{S}}\sigma_{i}^{x}\sigma_{j}^{x},\label{eq:-4}
\end{equation}
is an all-to-all interaction among qubits in $\mathcal{S}$,
with $\sigma_{j}^{x}=\sigma_{j}^{+}+\sigma_{j}^{-}$. The condition
$\delta\gg\eta\Omega$~\cite{Sackett2000} ensures that the qubits
interact with each other via second-order processes, populating the
COM mode only virtually and keeping it disentangled from the qubits.
This approach is insensitive to the temperature of the COM mode. However,
its rates are intrinsically lower than the rates of the resonant sideband
resources~\eqref{eq:}, $(\eta\Omega)^{2}/\delta\ll\eta\Omega$.

\textit{Analog Spin-model Resource} $-$ Recent experiment~\cite{Kokail2019}
showed that VQE in ion traps can also employ a programmable analog
quantum simulator as entangling resource operation. Such analog quantum
simulator is realized by using a bichromatic laser beam that off-resonantly
couples all the ions to all the transverse phonon modes, ultimately
implementing the XY Hamiltonian~\cite{PorrasCirac2004,FriisExperimentPorrasCirac}
\begin{equation}
H_{XY}=\sum_{i=1}^{N_{\text{ions}}-1}\sum_{j=i+1}^{N_{\text{ions}}}\frac{J_{ij}}{2}
\left(\sigma_{i}^{x}\sigma_{j}^{x}+\sigma_{i}^{y}\sigma_{j}^{y}\right)+B\sum_{i=1}^{N_{\text{ions}}}\sigma_{i}^{z}\label{eq:-6}
\end{equation}
with power-law $J_{ij}\sim J_{0}\left|i-j\right|^{-\alpha}$ long-range
couplings, where $\sigma_{j}^{\{x,y,z\}}$ are the Pauli operators
acting on qubit $j$ and $B$ is a uniform magnetic field. Thanks
to the off-resonant couplings, the interaction~\eqref{eq:-6} is
also insensitive to the temperature of the phonon modes. However, typical
rates $J_{0}\sim0.05-0.5\,\text{kHz}$~\cite{Smith2016,Elben2019}
are, again, much slower than the rates of resonant sideband interactions~\eqref{eq:}.

\section{Variational Quantum Eigensolvers}

\label{sec:VQExplained}

VQE enables the experimental realization of ground states of arbitrary
Hamiltonians, providing the best approximation for the given resource
operations. 
Once the optimization is converged, the optimal quantum state can
then be re-prepared on demand. For example, this state can be used
as a resource to initialize the system for digital quantum simulation~\cite{Georgescu2014},
which was recently demonstrated on ion traps in the context of lattice
gauge theories~\cite{Martinez2016}.

\textit{Target model: Su-Schrieffer-Heeger $-$} As an example, in
this work, we consider VQE for the ground state preparation of the
SSH model on ion qubits. The SSH is an exactly solvable Hamiltonian~\cite{SSHmodel1979,SSHmodel1983,AsbothSSH2016},
and a free-fermion prototype for symmetry-protected topological order
in 1D, which reads 
\begin{multline}
H_{\text{SSH}}=\sum_{j=1}^{N_{\text{ions}}-1}\left(1-\left(-1\right)^{j}t\right)\left(\sigma_{j}^{x}\sigma_{j+1}^{x}+\sigma_{j}^{y}\sigma_{j+1}^{y}\right)\\
+\tilde{B}\left(\sigma_{1}^{z}-\sigma_{N_{\text{ions}}}^{z}\right)\label{eq:-5}
\end{multline}
in open boundary conditions. 
$H_{\text{SSH}}$ is real in the canonical basis, exhibits $z$-magnetization
symmetry $[H_{\text{SSH}},\sum_{j}\sigma_{j}^{z}]=0$, and has zero-magnetization
ground states with real amplitudes. We added a small staggered magnetic
field $\tilde{B}=0.1$ at the boundaries, to lift the four-fold degeneracy
arising in the topological phase ($t<0$). With this prescription,
$H_{\text{SSH}}$ is also CP symmetric (invariant under global spin-flip plus spatial reflection), translationally invariant in the bulk
with a period of two sites, and has a critical point at $t=0$.

\textit{VQE with Closed System Analog resources (CSA) $-$} In Ref.~\cite{Kokail2019},
VQE was performed on trapped ions using Closed-System programmable
Analog resources as in Eq.~\eqref{eq:-6}. We report the variational
circuit ansatz considered in that work in Fig~\ref{fig:-9}(b). This
circuit alternates global resource operations, generated by $H_{XY}$,
and single-qubit rotations $\exp(-i\theta\sigma_{j}^{z})$. It was
shown that this strategy is potentially scalable and capable of approximating
ground states of 1D lattice models, including lattice gauge theories,
with high fidelity. Here, we consider the CSA circuit as a comparison
benchmark for the QDB-MPS circuit in preparing of the ground state of
$H_{\text{SSH}}$. We briefly review some details of the CSA strategy
to highlight the differences with respect to the QDB-MPS.

The variational CSA circuit built with closed-system resource interactions
$H_{k}^{(r)}\in\{H_{XY},\sigma_{j}^{z}\}$ generates (ideally pure) trial states
\begin{align}
\left|\psi_{\text{out}}(\boldsymbol{\theta})\right\rangle =\prod_{k}\text{exp}(-i\theta_{k}H_{k}^{(r)})\left|\psi_{\text{in}}\right\rangle,
\label{eq:-55}
\end{align}
where the product of operators is ordered right-to-left for increasing $k$.
The system qubits are initialized in a N{\'e}el
product state $\left|\psi_{\text{in}}\right\rangle =\left|\downarrow\right\rangle _{1}\left|\uparrow\right\rangle _{2}\left|\downarrow\right\rangle _{3}...\left|\uparrow\right\rangle _{N_{\text{ions}}}$
which can be prepared with high fidelity. For each given set of trial
parameters $\boldsymbol{\theta}$, multiple instances of $\left|\psi_{\text{out}}(\boldsymbol{\theta})\right\rangle $
are prepared, and projectively measured in various local bases to
reconstruct the energy cost function $\left\langle H_{\text{SSH}}\right\rangle _{\boldsymbol{\theta}}=\left\langle \psi_{\text{out}}(\boldsymbol{\theta})\right|H_{\text{SSH}}\left|\psi_{\text{out}}(\boldsymbol{\theta})\right\rangle $
from one- and two-point correlation functions $\langle\sigma_{i}^{a}\rangle$,
$\langle\sigma_{i}^{a}\sigma_{j}^{a}\rangle$. The cost function $\left\langle H_{\text{SSH}}\right\rangle _{\boldsymbol{\theta}}$,
function of $\boldsymbol{\theta}$, is evaluated at finite precision
and minimized with a search algorithm capable of taking into account
statistical errors. 
Upon convergence, $\left\langle H_{\text{SSH}}\right\rangle _{\boldsymbol{\theta}_{\text{opt}}}$
reaches its global minimum, and the optimal set of parameters $\boldsymbol{\theta}_{\text{opt}}$
can be used to re-prepare the (approximate) non-degenerate ground
state $\left|\psi_{\text{out}}(\text{\ensuremath{\boldsymbol{\theta}_{\text{opt}}}})\right\rangle $
on demand.

Convergence of the search algorithm 
can be enhanced by incorporating symmetries of the target model in
the trial states. For the CSA circuit, the zero $z$-magnetization
of the initial N{\'e}el state $\left|\psi_{\text{in}}\right\rangle $
is protected both by the entangler $H_{XY}$ and rotations $\text{exp}(-i\theta_{j}\sigma_{j}^{z})$.
CP symmetry is protected by $H_{XY}$, and can be enforced on single-qubit
transformation `layers' by using correlated rotations angles $\theta_{j}=-\theta_{N_{\text{ions}}+1-j}$.
can be approximated as well, by repeating odd-site and even-site rotation
angles, $\theta_{j}=\theta_{j+2}$, away from the system edges~\cite{Kokail2019}.

While VQE provides the best approximate state preparation for a given
set of resources in the presence of actual imperfections and noise~\cite{RaphaelSSS},
it is still affected by decoherence. By considering a different VQE
circuit ansatz for trapped ions, which uses the high-rate resonant
sideband resources of Eq.~\eqref{eq:}, we aim to tackle decoherence
by speeding-up the preparation of trial states.

\textit{VQE with Quantum Data Buses (QDB-MPS)$-$} We design a variational
circuit ansatz $U(\boldsymbol{\theta})=\prod_{k}\text{exp}(-i\theta_{k}H_{j_k})$
built on ion-COM mode interaction pulses~\eqref{eq:},
according to Fig.~\ref{fig:-1}(b). Since now the COM mode becomes
entangled to the ion qubits, the state of the qubits is generally
mixed, both during the processing and in the output trial state 
\begin{align}
\rho_{\text{out}}(\boldsymbol{\theta})=\text{Tr}_{\text{COM}}[U(\boldsymbol{\theta})(\left|\psi_{\text{in}}\right\rangle \left\langle \psi_{\text{in}}\right|\otimes\rho_{0})\,U^{\dagger}(\boldsymbol{\theta})]. \label{eq:-56}
\end{align}
Parameters $\{\theta_{k}\}$ can be controlled by the duration of the 
sideband laser pulses. Similar to the closed-system case, the cost
function $\langle H_{\text{SSH}}\rangle_{\boldsymbol{\theta}}=\text{Tr}[\rho_{\text{out}}(\boldsymbol{\theta})H_{\text{SSH}}]$
is reconstructed from correlation functions evaluated on the trial
qubit state, while the output state of the COM mode is discarded.
As $\langle H_{\text{SSH}}\rangle_{\boldsymbol{\theta}}$ is minimized
to approach the ground energy, $\rho_{\text{out}}(\boldsymbol{\theta})$
approximates the unique, pure ground state $|\psi_{\text{targ}}\rangle\langle\psi_{\text{targ}}|$,
and the output state of the COM mode becomes automatically disentangled.

In our design, resource operations are arranged such that the circuit
ultimately generates a variational Matrix Product State (MPS). In
fact, MPS efficiently approximate ground states of 1D quantum lattice
Hamiltonians, including the SSH model~\cite{Schollwock2011,Orus2014}.
In contrast to CSA resources, which are inherently global, blue-sideband
operations~(\ref{eq:}) offer single-site addressability. This property
allows us to progressively tailor the MPS from one edge to the other,
as shown by the Tensor Network diagram~\cite{Schollwock2011,Orus2014}
in Fig.~\ref{fig:-1}(c), in analogy to quantum circuits developed
in the context of quantum machine learning~\cite{Huggins2018}.
Previous strategies for realizing MPS on quantum devices were proposed,
both for deterministic state preparation~\cite{Schon2007,Ran2019},
as well as for variational state preparation~\cite{Liu2019} which
employed long-range interactions. In contrast, our variational circuit
requires only local ion-phonon interactions, and, therefore, can be
up-scaled to modular ion trap architectures (see next section). Additionally,
we introduce an approximate bulk-translational invariance to the ansatz,
to reduce its variational complexity: In our circuit, interactions
in the bulk can be virtually grouped in `boxes' (drawn in Fig.~\ref{fig:-1}(b)
as grey rectangles at the edges, and orange and green for the bulk).
Each box is defined by a set of variational parameters, which are
repeated in the bulk boxes of the same color.
At the same time, the maximum achievable bond dimension $D\sim2^{l-1}$
of the generated MPS can be controlled by increasing the size of the
bulk boxes $l$ (see Appendix~\ref{sec:Maximal-bond-dimension}).
Figure \ref{fig:-1} shows a specific design of the bulk boxes using five operations per box.
Numerical results suggest that the number of operations per box is crucial towards the accuracy of the
ansatz, while the ordering is marginal (data not shown).
Appendix \ref{sec:Gates-sequence-in} illustrates our basic strategy to arrange operations
inside a box.

As in the CSA case, we again incorporate symmetries of the target
model in the trial states. First of all, since~$H_{j}$ is fully
imaginary, the operation $U(\boldsymbol{\theta})$ is real as well
as $\left|\psi_{\text{in}}\right\rangle $ and $\rho_{0}(T)$, and,
thus, the generated states $\rho_{\text{out}}(\boldsymbol{\theta})$
have real amplitudes.
Moreover, operations~(\ref{eq:}) protect an `extended' magnetization
symmetry $1/2\sum_{j}\sigma_{j}^{z}-a^{\dagger}a$, forcing the output
qubits state to have well-defined $z$-magnetization upon qubits and
QDB becoming variationally disentangled (see Appendix~\ref{sec:Symmetries}
for details). We have analytically verified that, in the theoretical
limit of infinite-depth circuits, the resources given by Eq.~(\ref{eq:})
are sufficient to generate ground states of \textit{\emph{any}} real,
magnetization conserving Hamiltonian (proof in Appendix~\ref{sec:Controllability}).
Controllability for an arbitrary lattice model (e.g., without symmetries)
can be achieved by adding to the resources a resonant carrier operation
$\sigma_{j}^{z}$ for at least one ion.

We stress that resources of the form ~\eqref{eq:} are typical in
various experimental platforms with bosonic QDBs, such as superconducting
qubits connected via a resonator~\cite{Sillanpaa2007,Fink2017,Song2017}.
As such, the numerical analysis presented below can be generalized
to these platforms.

\textit{CSD-MPS Circuit Ansatz $-$} We also compare our QDB-based
approach with a variational MPS circuit ansatz realized with Closed
System Digital resource operations (CSD-MPS circuit), as presented
in Fig~\ref{fig:-9}(c). This circuit is built with single-qubit
$\sigma_{i}^{z}$ rotations and site-filtered MS gates~\eqref{eq:-4}
applied to a local set of ions. Such resource operations protect z-magnetization
in the trial state.

\section{\textit{\emph{Results}}}

\label{sec:results}

\textit{Performance$-$ }We compare the performance of the variational
circuits by numerically simulating VQE for preparation of the ground
state $|\psi_{\text{targ}}\rangle$ of Hamiltonian~\eqref{eq:-5}
in the gapped phase with $t=0.5$. Trial states are obtained by evolving
the initial quantum state in terms of density matrixes. For the QDB-MPS
circuit, we consider the COM mode to be prepared in a thermal state
$\rho_{0}(T)$ with realistic average number of excitations $n_{0}(T)=\text{Tr}[a^{\dagger}a\,\rho_{0}(T)]=1/(e^{1/T}-1)=0.01$~\cite{Roos1999,Schafer2018}.
Other sources of imperfections are not taken into account, and, therefore,
in contrast to the QDB-MPS circuit, simulations of the CSA and CSD-MPS
circuits are free of imperfections. We consider the theoretical limit
of unlimited measurement budget~\cite{RaphaelSSS,Kokail2019}, i.e.,
the cost function is the \textit{exact} expectation value $\langle H_{\text{SSH}}\rangle_{\boldsymbol{\theta}}$.
To optimize parameters $\boldsymbol{\theta}$, we use a gradient-based
optimization algorithm (see Appendix~\ref{sec:Parameters-optimization}),
which can be adapted for noisy cost functions~\cite{Borkar2015,A.2015,Leng2019}.
We emphasize that we consider a low number of free variational parameters,
$N_{p}\sim10-25$, which enables global search methods, such as the
dividing rectangles algorithm~\cite{NicholasDIRECT,LiuDIRECT},
already adopted in VQE~\cite{Kokail2019}. For the analog resource
entangler $H_{XY}$, realistic values $\alpha=1.34$ and $B=20$~\cite{Kokail2019}
are used. Further details on the circuit ans{\"a}tze and their simulation
are given in Appendix~\ref{sec:Gates-sequence-in}.

\begin{figure}
\includegraphics[width=\linewidth]{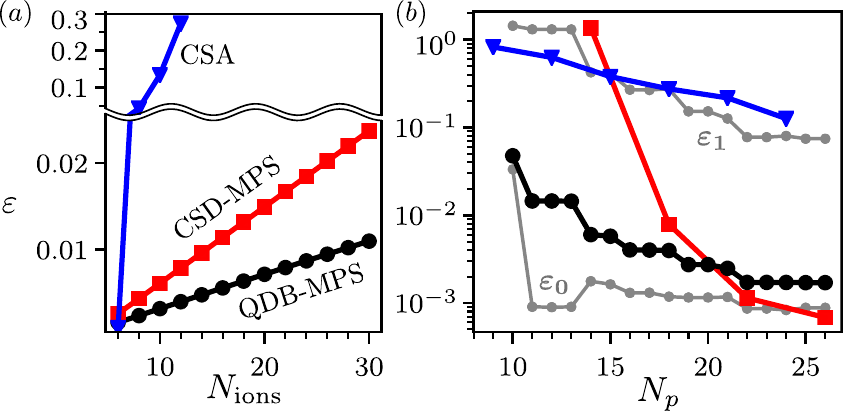}\caption{Energy error related to the excitation gap, $\varepsilon$, (a) versus
size of the prepared states $N_{\text{ions}}$ for fixed number variational
parameters $N_{p}=18$ and (b) versus $N_{p}$ for $N_{\text{ions}}=12$
in circuits QDB-MPS, CSD-MPS, and CSA, as indicated in (a). Gray small rounds in (b) give values of $\varepsilon_{q}$ obtained in the same QDB-MPS
circuit and parameters as the black plot but with different initial
pure states of the QDB $\left|q\right\rangle _{a}$ for $q\in\{0,1\}$.}
\label{fig:excitationenergy}
\end{figure}

To assess the accuracy of the ground state preparation, we employ
various figures of merit. Besides the fidelity
and the excitation energy (defined later on), we also consider a correlation-based
error function
\begin{align}
F_{\text{err}} & =\sum_{i=1,\,j>i}^{N_{\text{ions}}}\left|C_{ij}^{\text{DMRG}}-C_{ij}\right|/\sum_{i=1,\,j>i}^{N_{\text{ions}}}\left|C_{ij}^{\text{DMRG}}\right|,\label{eq:-3}
\end{align}
calculated using (parallel) two-point correlators 
\begin{align}
C_{ij} & =\langle\sigma_{i}^{x}\sigma_{j}^{x}\rangle-\langle\sigma_{i}^{x}\rangle\langle\sigma_{j}^{x}\rangle.\label{eq:-1}
\end{align}
Here $C_{ij}$ is evaluated on the optimized state $\rho_{\text{out}}\left(\boldsymbol{\theta}_{\text{opt}}\right)$
and $C_{ij}^{\text{DMRG}}$ is the exact correlator of $|\psi_{\text{targ}}\rangle$
obtained by DMRG method \cite{DMRGWhite}. We consider this error
function in Fig.~\ref{fig:-9}(a), to show that the proposed QDB-MPS
circuit is able to prepare states with $F_{\text{err}}$ saturating
in $N_{\text{ions}}$ at fixed number of variational parameters $N_{p}$.
In particular, the insets show that these states have correlations
$C_{ij}$ accurately capturing $C_{ij}^{\text{DMRG}}$. By contrast,
$F_{\text{err}}$ arising from the CSA circuit grows rapidly for this
specific task, and the correlators $C_{ij}$ decay slower than in
$|\psi_{\text{targ}}\rangle$. We argue that this follows from the
necessity to variationally remove the strong (power-law) correlations
established by the long-range interactions in $H_{XY}$~\eqref{eq:-6}
to recover the exponential decay of correlations in the ground state.
Moreover, the SSH ground state exhibits considerable (short-range)
entanglement dimerization, which in turn seems to favor quasi-local
resource circuits over global resource circuits. While the results
for CSD-MPS and QDB-MPS circuits given in Fig~\ref{fig:-9}(a) are
compatible, the QDB-MPS is realized with higher-rate operations, and,
therefore, is expected to be less prone to decoherence.

\textit{\emph{S}}\emph{calability}\textit{ $-$} We study the scalability
of the circuits in terms of the excitation energy in units of the
energy gap, which bounds infidelity of the prepared states $1-F\le\varepsilon=(\langle H_{\text{SSH}}\rangle_{\boldsymbol{\theta}_{\text{opt}}}-E_{0})/\Delta$~(Appendix
\ref{sec:Energy-bounds-on}), and is thus a figure of accuracy.
Here, $F=\langle\psi_{\text{targ}}|\rho_{\text{out}}\left(\boldsymbol{\theta}_{\text{opt}}\right)|\psi_{\text{targ}}\rangle$
and $\Delta=E_{1}-E_{0}$ is the energy gap obtained by DMRG, which
converges to a finite non-zero value in the thermodynamical limit
$N_{\text{ions}}\to\infty$ since the target model is non-critical.
Figure~\ref{fig:excitationenergy}(a) shows that, for the QDB-MPS and
CSD-MPS circuits, $\varepsilon$ grows linearly with $N_{\text{ions}}$, in
contrast to the CSA circuit. This is compatible with the previous
observation that the two-point correlation functions carry, on average,
an error saturating in $N_{\text{ions}}$, and the energy functional
contains a linear amount of such correlators.

Scaling of the accuracy $\varepsilon$ with $N_{p}$, for the various
methods and at fixed $N_{\text{ions}}$, is plotted in Fig.~\ref{fig:excitationenergy}(b).
Again, we observe that MPS circuits offer asymptotically better accuracy
(or better parametric efficiency) than the CSA global resources.
While increasing the number of parameters $N_{p}$, we observe the
CSD-MPS circuit reaches slightly better accuracies than the QDB-MPS
circuit; we attribute this effect to the finite temperature of the
COM mode we employed for the simulations.
This argument is supported by the gray curves in Fig.~\ref{fig:excitationenergy}(b),
showing relative excitation energies $\varepsilon_{q}$ obtained with
the same QDB-MPS circuit and the same parameter values as the black
plot but with different initial pure states of the COM mode $\left|q\right\rangle _{a}$
for $q\in\{0,1\}$. For $n_{0}=0.01$, we have $\varepsilon=\varepsilon_{0}p_{0}+\varepsilon_{1}p_{1}+O(n_{0}^{2})$,
with $p_{q}=\langle q|_{a}\rho_{0}(T)|q\rangle_{a}\sim n_{0}^{q}$.
The plot shows that the two quantum trajectories are both variationally
optimized in such a way that their error contributions are commensurate,
$\varepsilon_{0}p_{0}\sim\varepsilon_{1}p_{1}$.

\begin{figure}
\includegraphics[width=\linewidth]{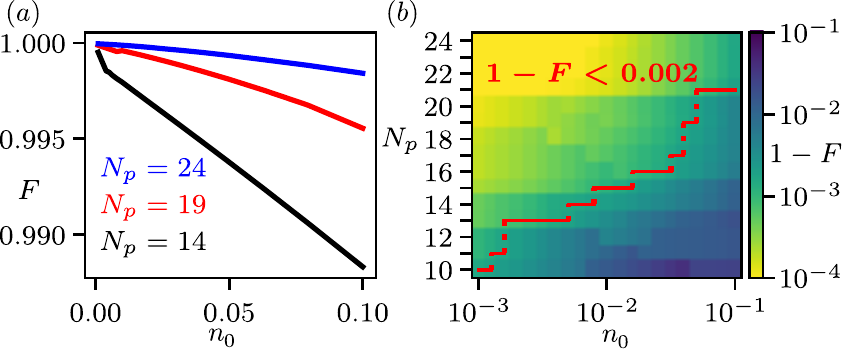}

\caption{(a) Fidelity and (b) infidelity of the $6$ qubit state prepared in
the QDB-MPS circuit as functions of number of excitations $n_{0}$
in $\rho_{0}(T)$ and number of variational
parameters $N_{p}$ in the circuit. In (b), the red line indicates
the lowest number of $N_{p}$ required to preserve infidelity threshold
$1-F\le0.002$ at given $n_{0}$.}

\label{fig:temperature}
\end{figure}

\textit{Tolerance to temperature $-$} Here, we further investigate
tolerance of the QDB-MPS approach to finite temperatures $T$ of the
COM mode. As $T$ increases, more amplitudes $p_{q}$ are populated
in $\rho_{0}(T)=\sum p_{q}(T)\left|q\right\rangle _{a}\left\langle q\right|$,
and, therefore, more equations $U\left(\boldsymbol{\theta}_{\text{opt}}\right)\left|\psi_{\text{in}}\right\rangle \left|q\right\rangle _{a}\sim\left|\psi_{\text{targ}}\right\rangle \left|q\right\rangle _{a}$
need to be simultaneously satisfied, requiring higher $N_{p}$ to
achieve the same accuracies. In Fig.~\ref{fig:temperature}, we consider
the fidelity of the optimized output state of $N_{\text{ions}}=6$
qubits, as a function of the average number of excitations $n_{0}(T)$
and number of variational parameters $N_{p}$. Panel (a) shows that,
for fixed $N_{p}$, $F$ decreases linearly with $n_{0}$ (similarly
to the C-phase gate~\cite{Cirac1995} built with sidebands), but
with a slope that flattens for increasing $N_{p}$. Therefore, to
achieve a fidelity threshold, $N_{p}$ must be increased with $n_{0}$.
Remarkably, for the case under consideration, it seems that the $N_{p}$
required to obtain a static fidelity threshold scales logarithmically
with $n_{0}$. This behavior is illustrated in panel (b), where the
density plot shows the infidelity of the optimized state as a function
of $n_{0}$ and $N_{p}$. The red curve highlights a fixed fidelity
threshold $1-F\le0.002$.
This property sets our QDB-based VQE apart from the approaches using
sideband interactions to construct logical gates~\cite{Cirac1995},
in which imperfect cooling imposes a hard bound on the attainable
accuracies.

\begin{figure}
\includegraphics[width=\linewidth]{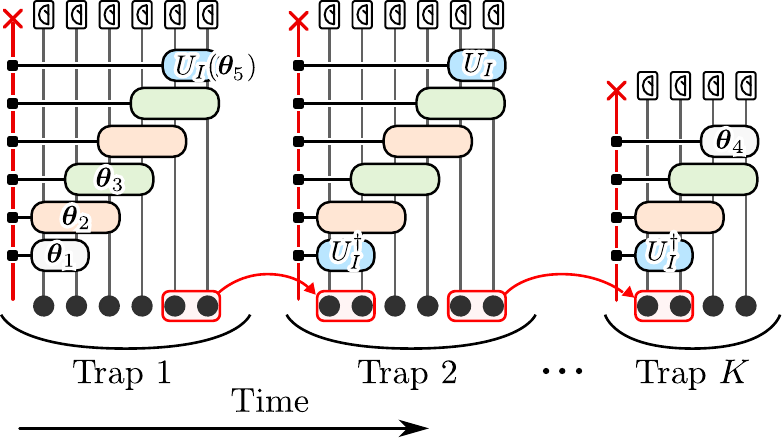}

\caption{QDB-MPS circuit in a modular ion trap architecture, which generates
a state on the ions chain with 4 ions per trap. The processing is
implemented consequently, starting from the left trap to the right,
and from the bottom to the top in a single trap as presented by the
circuits. The red boxes with the arrows indicate the transfer of the
edge ions (or their internal states) to the next trap after the implementation of all operations
in a trap.}
\label{fig:-5}
\end{figure}

\textit{Scalability on a }\emph{modular ion trap architectures}\textit{
$-$} The proposed QDB-MPS circuit~\ref{fig:-1}(a) relies on interactions
between individual ions and the COM mode. However, the number of ions
which can be coupled to the COM mode in a single trap is limited by
$\sim10-100$~\cite{Monroe2013}. To further scale the ion trap
quantum processors, modular ion trap architectures were proposed~\cite{Blatt2008,Monroe2013}.
Following these proposals, we describe ground state preparation with
QDB-MPS circuit across multiple traps, as shown in Fig.~\ref{fig:-5}.
We consider coherent shuttling of few ions~\cite{Blatt2008,Bowler2012}
(or their internal states~\cite{Monroe2013}) as a quantum communication
channel among the traps, and we assume it to be error-free. We consider
each trap containing a small number of ions, such that each ion is
addressable via blue sideband pulses, realizing interactions as Eq.~\eqref{eq:}
with the COM mode of the corresponding trap. For simplicity, we assume
that the initial thermal states, $\rho_{0}$, of the COM modes are
identical in each trap, as well as and their effective coupling to
the ions.

The variational circuit ansatz is constructed sequentially along the
network of traps, i.e., by applying operations in trap $k+1$ after
all operations in trap $k$ are completed. Since each trap has an
independent COM mode as a bosonic QDB, we introduce a variational
`interface' operation $U_{I}(\boldsymbol{\theta}_{5})$, constructed
with resource interactions~\eqref{eq:}, which is applied to a set
of $l-1$ ions (with $l$ the size of the bulk boxes) as the last
operation for trap $k$. This operation can be optimized in order
to disentangle the COM mode of trap $k$, effectively implementing
$U_{I}\rho_{\text{qubits},k}U_{I}^{\dagger}=\rho_{\text{qubits}}\otimes\rho_{0}$,
with $\rho_{\text{qubits},k}$ the common state of ions and the COM
mode.
 After transferring the set of $l-1$ ions to trap $k+1$, we apply
the unitary $U_{I}^{\dagger}$ to the same ions plus the COM mode
of trap $k+1$ prepared in state $\rho_{0}$, effectively reconstructing
the state $\rho_{\text{qubits},{k+1}}=U_{I}^{\dagger}(\rho{}_{\text{qubits}}\otimes\rho_{0})U_{I}=\rho_{\text{qubits},k}$.
Ultimately, this operation removed the `impurity' introduced by the
interface between the two traps.

\begin{figure}
\includegraphics[width=\linewidth]{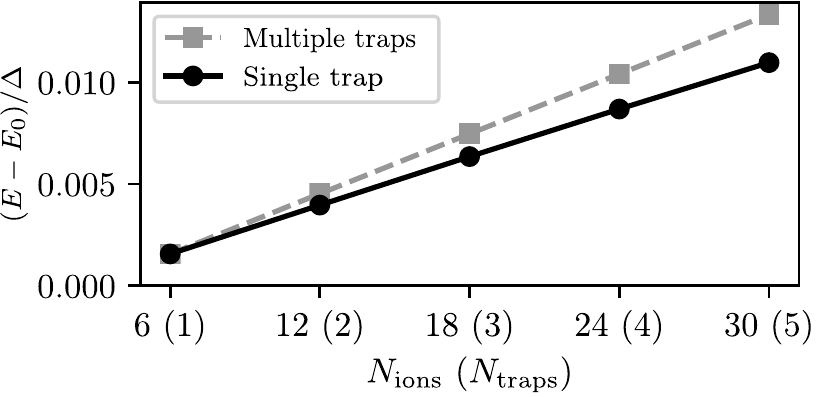}\caption{Energy error related to the excitation gap versus size $N_{\text{ions}}$
of the optimized state in a single trap using $N_{p}=18$ variational
parameters and in $N_{\text{traps}}$ traps with 6 ions per trap using
$N_{p}=18+2$ .}
\label{fig:-6}
\end{figure}

In Fig.~\ref{fig:-6}, we compare the excitation energies, from optimizing
the QDB-MPS circuit, with a single trap and with a modular trap network.
Both circuits have the same number of variational operations in the
boxes in the edges and in the boxes of the bulk, and the interface operation $U_{I}$
contains $2$ variational sideband operations. The error emerging
from the optimization of the ion-trap network deviates only slightly from the
error in the single trap case, and we recover linear scaling of the
error with the system size. This result is an even stronger indicator
of scalability of the QDB-MPS approach for VQE, as it can treat hardware
interfaces as impurities, and efficiently compensate for them with
small variations of the circuit ansatz.

\section{Outlook}

\label{sec:outlook}

We proposed an approach for VQE which uses qubit-QDB interactions,
available in currently-existing programmable quantum devices, as variational
resource operations. In our strategy, the QDB is disentangled from
the optimized states only at the end of the state preparation, as
a byproduct of the optimization. For trapped ions, we discussed how
to realize our strategy with resonant blue-sideband pulses, potentially
yielding faster state-preparation rates compared to existing strategies,
and thus helping to mitigate decoherence effects. The variational
ansatz circuit we designed is tailored to efficiently prepare MPS,
leading to a potential improvement in scalability over current VQE strategies.

We numerically simulated the ion trap implementation of VQE based
on the QDB-MPS ansatz circuit, and showed strong evidence of scalability
of our approach, even when it is realized in modular ion traps. While
the ions-COM mode sideband interactions operate in an extended Hilbert
space, we demonstrated that these resource operations can incorporate
symmetries in the variational circuit, similar to the closed-system
VQE approaches. Although here we demonstrated robustness of the method
to the initial temperature of the COM mode, further investigations
are required to study other realistic imperfections, including (i)
qubits decoherence, (ii) fluctuations of the control parameters, and
(iii) shot noise resulting from a finite budget of measurements to
reconstruct the cost function.

The developed QDB-based VQE can be readily used in other currently
available experimental platforms, such as quantum dots~\cite{Baart2017}
or atoms coupled by (chiral) waveguides~\cite{Lodahl2016}. Moreover,
we expect that the scaling and parametric efficiency of our results,
found for the MPS circuit ansatz, extend to other efficiently-contractible
tensor network states~\cite{Huggins2018,Liu2019}, such as $\text{MPS}{}^{2}$~\cite{Zaletel2019,Haghshenas2019},
thus allowing us to accomodate the entanglement content required to explore higher dimensions.
Also, by including simultaneous operations on multiple qubits, it is possible
to build many-body wavefunction ans\"atze beyond tensor networks, ultimately enabling the
efficient preparation of quantum states that can not be numerically simulated.
Finally, we envision
the possibility of including controllable dissipative elements to
the QDB resources \cite{Rivas2012} in order to construct optimized cooling algorithms
beyond stabilizer codes~\cite{StabilizerShor,StabilizerFlorentin}.
\begin{acknowledgments}
We warmly thank M. Hettrich, R. Kaubruegger, C. Kokail, M. Meth, T.
Monz, P. Schindler R. van Bijnen, and P. Zoller for private discussions.
Authors kindly acknowledge support by the EU Quantum Technology Flagship
project PASQUANS and the QuantEra project QTFLAG, the Austrian Research
Promotion Agency (FFG) via QFTE project AutomatiQ,
the US Air Force Office of Scientific Research (AFOSR)  via  IOE  Grant  No.  FA9550-19-1-7044 LASCEM,
and
by the Army Research Laboratory and was accomplished under Cooperative
Agreement Number W911NF-15-2-0060. The views and conclusions contained
in this document are those of the authors and should not be interpreted
as representing the official policies, either expressed or implied,
of the Army Research Laboratory or the U.S. Government. The U.S. Government
is authorized to reproduce and distribute reprints for Government
purposes notwithstanding any copyright notation herein. The computational
results presented have been achieved (in part) using the HPC infrastructure
LEO of the University of Innsbruck. Source codes used for our results are available
at \url{https://github.com/Viacheslav-Kuzmin/quantum_bus_vqe},
released under a MIT license.
\end{acknowledgments}

\newpage{}

\section*{Appendix}

\appendix
%dummy comment 

\section{Maximum bond dimension\label{sec:Maximal-bond-dimension}}

\begin{figure}
\includegraphics[width=\linewidth]{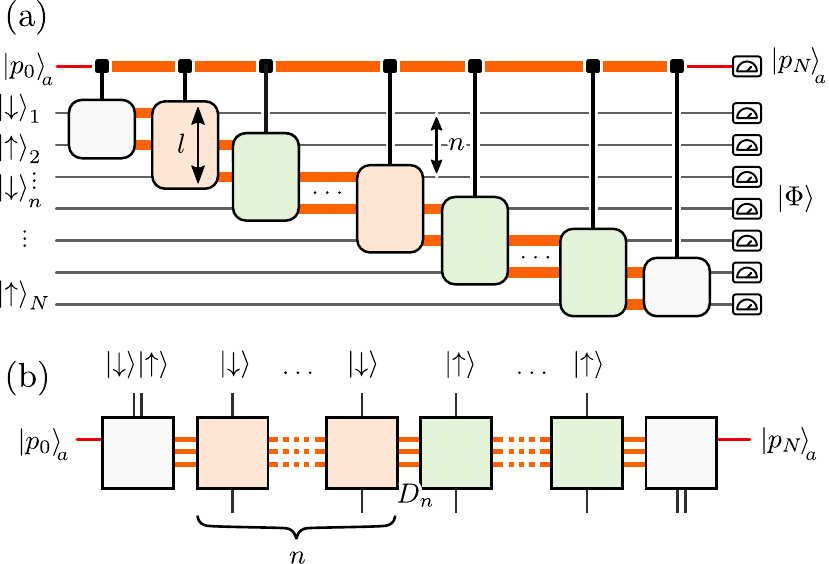}

\caption{(a) QDB-MPS circuit with the measurement of the bosonic QDB, which results
in probabilistic generation of  MPS $\left|\Phi\right\rangle $ in
qubits. (b) Tensor network diagram of $\left|\Phi\right\rangle $.}
\label{fig:-8}
\end{figure}

In the section, we study the maximum bond dimension, $D_{M}$, of
MPS generated with the QDB-MPS circuit using a bosonic QDB and built with resource operations~\eqref{eq:} described by the Anti-Jaynes-Cummings Hamiltonian. Here we distinguish two sets of MPS accessible by
the circuit. The first set contains all pure states which can be generated
by the QDB-MPS circuit. The states of this set can be generated probabilistically~\cite{Schon2007}
after measuring (and therefore disentangling) the QDB at the end.
It defines the search space for VQE and, therefore, is desired to
be limited. The second set --- a subset of the first set --- contains
deterministically generated translationally invariant MPS, which are
disentangled from the QDB by the end of the circuit without measuring
the EM. This set, in opposite to the first one, is required to be
big enough to include a good approximation of the target state.

\subsection*{Set of probabilistically generated MPS}

We start with the set of MPS generated probabilistically. We characterize
this set of MPS by considering a possible range of their bond dimensions.
Bond dimension $D_{M}$ of MPS of size $N$ is the maximum of bond
dimensions $D_{n}$ between two blocks of qubits $[1,n]$ and $[n+1,N]$.
$D_{n}$ corresponds to entanglement entropy between the blocks~\cite{Orus2014},
such that the Schmidt decomposition of MPS $\left|\Phi\right\rangle $
in the basis of two blocks $\{|\phi_{[1,n]}^{(i)}\rangle\}$ and $\{|\phi_{[n+1,N]}^{(i)}\rangle\}$
is 
\[
\left|\Phi\right\rangle =\sum_{i=1}^{D_{n}}a_{i}|\phi_{[1,n]}^{(i)}\rangle|\phi_{[n+1,N]}^{(i)}\rangle.
\]
The entanglement entropy of the blocks is $S_{[1,n]}=-\text{Tr}\left[\rho_{[1,n]}\text{log}(\rho_{[1,n]})\right]\le\text{log}D_{n}$,
with $\rho_{[1,n]}$ the state of qubits $[1,n]$.

Using singular value decomposition, one can show that, for an arbitrary
state of size $N$, it is valid that $D_{n}\le2^{\text{min}(n,N+1-n)}$,
and, therefore, $D_{M}\le2^{\left\lfloor N/2\right\rfloor }$. By
using a not universal set of resource operations~\eqref{eq:-2} we
limit $D_{M}$ and, therefore, restrict the set of achievable states.
As presented in Appendix~\ref{sec:Symmetries}, interactions~\eqref{eq:}
preserve an extended symmetry $Z=1/2\sum_{j=1}^{N}\sigma_{j}^{z}-a^{\dagger}a$.
In the following, we explain that this symmetry imposes $D_{n}\lessapprox\left\lfloor n/2\right\rfloor \cdot2^{l-1}$,
where $l$ is the size of the bulk boxes in the QDB-MPS circuit, and
limits the maximum bond dimension as $D_{M}\lessapprox\left\lfloor 2/3\cdot N\right\rfloor 2^{l-1}$.
Therefore, in comparison with the general case, the bond dimension
of the MPS generated by the QDB-MPS circuit is exponentially suppressed.

Figure~\ref{fig:-8}, demonstrates entanglement distribution in the
QDB-MPS circuit between the qubits by a subsystem, highlighted by
orange color, which includes the QDB and $l-1$ qubits shared by neighbor
virtual boxes. Such a subsystem can be considered
as a virtual ancilla that sequentially interacts with the qubits distributing
entanglement among them~\cite{Schon2007}. The effective dimension
of this ancilla (dimension of the populated Hilbert space) bounds
bond dimension $D_{n}$.

Let us consider the QDB initially prepared in state $\left|p_{0}\right\rangle _{a}=\left|0\right\rangle _{a}$,
and the qubits prepared in state $\left|\psi_{\text{in}}\right\rangle =\left|\downarrow\right\rangle _{1}\left|\uparrow\right\rangle _{2}\left|\downarrow\right\rangle _{3}...\left|\uparrow\right\rangle _{N}$.
Then, because of the preserved symmetry, the maximum possible population
of the QDB after interaction with $n$ ions can not be higher than
$\left|\left\lfloor n/2\right\rfloor \right\rangle _{a}$. Multiplying
the growing dimension of the QDB by the dimension of $l-1$ qubits
in the ancilla, we obtain that $D_{n}\lessapprox\left\lfloor n/2\right\rfloor \cdot2^{l-1}$.
From the other side, as explained in~\cite{Schon2007}, $D_{n}$
has to decrease as $n$ approaches to $N$ to guarantee that the ancilla
(with QDB measured in $\left|p_{N}\right\rangle _{a}$) is decoupled by the
end of the state preparation. Therefore, there exists some $n_{M}$,
for which $D_{n_{M}}$ achieves its maximum. To obtain $n_{M}$, we
study how $D_{n}$ can decrease while the ancilla interacts with the
qubits after qubit $n_{M}$. Intuitively, this can be done other way
around by considering the growth of bond dimension $\tilde{D}_{N-n}$
if MPS would be generated from the other side of the qubits chain,
starting from $n=N$. Since the measured output state $\left|p_{N}\right\rangle _{a}$
can contain number of excitations $p_{N}\ge0$, the number of excitations
in the QDB can not only increase but also decrease as the QDB interacts
with the qubits, populating Hilbert space with the twice bigger dimension
than if the QDB would be initially prepared in $\left|0\right\rangle _{a}$.
Therefore, $\tilde{D}_{N-n}\approx2D_{\tilde{n}}\big|_{\tilde{n}=N-n}\lessapprox(N-n)2^{l-1}$.
Solving $D_{n_{M}}=\tilde{D}_{N-n_{M}}$ we obtain that, after $n_{M}\lessapprox\left\lfloor 2/3\cdot N\right\rfloor $,
$D_{n}$ has to decrease with $n$. Thus, we find that $D_{M}=D_{n_{M}}\lessapprox\left\lfloor 2/3\cdot N\right\rfloor 2^{l-1}$.

\subsection*{Set of deterministically generated translationally invariant MPS}

In order to generate MPS with translationally invariant $D_{n}$ in
the bulk, the virtual ancilla has to own dimension which repeats itself
with a target period along the state preparation. This includes the
ancilla dimension after realization of the last box in the bulk, after
which the edge box is aimed to disentangle the QDB from $l-1$ ancilla
qubits as required by deterministic preparation of a pure state. This
can be done only if the effective dimension of the ancilla does not
exceed dimension $2^{l-1}$ of the qubits. Hence, the limit of the
bond dimension of the translationally invariant MPS deterministically
generated in the considered QDB-MPS circuit is $\sim2^{l}$. Here,
we defined the limit not exactly, since the actual bond dimension
can vary within one period of the bulk.

\section{Symmetries\label{sec:Symmetries}}

In this section, we discuss the symmetries protected by the operations

\begin{equation}
H_{j}=i\tilde{\Omega}\left(a\sigma_{j}^{-}-a^{\dagger}\sigma_{j}^{+}\right),\label{eq:-2}
\end{equation}
with $\tilde{\Omega}$ the Rabi frequency, $a$ is the destruction
operator for the bosonic QDB, and $\sigma_{j}^{-}=(\sigma_{j}^{+})^{\dagger}=\left|\downarrow\right\rangle \left\langle \uparrow\right|_{j}$
the lowering operator for qubit $j$, with $\{\left|\downarrow\right\rangle _{j},\left|\uparrow\right\rangle _{j}\}$
the internal logical states of the qubit.

\textit{Magnetization} $-$ A number symmetry, forming an Abelian
$U(1)$ group of symmetric transformations $W(\varphi)=\text{exp}(-i\varphi Z)$,
is generated by the integer operator 
\begin{equation}
Z=1/2\sum_{j=1}^{N}\sigma_{j}^{z}-a^{\dagger}a=M-a^{\dagger}a,\label{eq:symmetryextend}
\end{equation}
where $N$ is the number of qubits, $M=1/2\sum_{j}\sigma_{j}^{z}$,
and $\sigma_{j}^{z}=\left|\uparrow\right\rangle \left\langle \uparrow\right|_{j}-\left|\downarrow\right\rangle \left\langle \downarrow\right|_{j}$.
It is straightforward to check that $Z$ is a symmetry since $[H_{j},Z]=0$
for every site $j\in\{1..N\}$. $Z$ thus defines a quantum number,
which is conserved during the controlled variational dynamics. Consider
the initial state of the system of qubits plus the QDB as $\left|\Psi_{\text{in}}\right\rangle =\left|\psi_{\text{in}}\right\rangle \otimes|s\rangle_{a}$,
with $\left|\psi_{\text{in}}\right\rangle =\left|\downarrow\right\rangle _{1}\left|\uparrow\right\rangle _{2}\left|\downarrow\right\rangle _{3}...\left|\uparrow\right\rangle _{N}$
the state of the qubits and $|s\rangle_{a}$ the Fock state of the
QDB, $a^{\dagger}a|s\rangle_{a}=s|s\rangle_{a}$, $s\in\mathbb{N}$.
This state has well-defined quantum number $Z\left|\Psi_{\text{in}}\right\rangle =s\left|\Psi_{\text{in}}\right\rangle $.
Due to the symmetry of Eq.~\eqref{eq:symmetryextend}, the evolution
of this state will preserve its quantum number, and the state will
always be of the form $\left|\Psi_{\text{out}}\right\rangle =\sum_{q,m,t}\delta_{m,q-s}c_{m,t}|m,t_{m}\rangle\otimes|q\rangle_{a}$,
where the qubit states are labeled by the magnetization $m$ (eigenvalues
of $M$) and degeneracy $t_{m}$. At the end of the variational quantum
state preparation, after optimization of the process, the output state
will be disentangled between phonon and qubits, and reads $\left|\Psi_{\text{opt}}\right\rangle =\left(\sum_{t}c_{m,t}|m,t_{m}\rangle\right)\otimes|m+s\rangle_{a}$.
The final qubit state will thus have a well-defined magnetization
$m\in\mathbb{N}$. This requirement identifies the class of target
Hamiltonians $H$ targetable with the chosen resources: which must
contain the magnetization symmetry: $[H,M^{z}]=0$.

\textit{Conjugation} $-$ A second symmetry is nested within our controls,
and it becomes evident after gauging the resource Hamiltonians as
in Eq.~(\ref{eq:}). In this format, the resource Hamiltonians are
purely imaginary $H_{j}=-\bar{H}_{j}$ when expressed in the canonical
basis, and thus antisymmetric $H_{j}=-H_{j}^{t}$ and traceless $\text{Tr}[H_{j}]=0$.
It follows that the unitary operators $U_{j}(t)=e^{-iH_{j}t/\hbar}$
of time evolution under $H_{j}$ must be real, as they are imaginary
exponentials of imaginary matrices, and have determinant $\text{det}(U_{j})=e^{-it\text{Tr}[H_{j}]}=1$.
This means that, with the chosen resources, we can only perform transformations
in the special orthogonal group SO($2^{d}$), a subgroup of the unitaries
U($2^{d}$), which, in turn, limits the set of states we can prepare
to real states (vectors with real coefficients in the canonical basis).
In conclusion, the class of models targetable with our variational
scheme is restricted to models expressed by real, or symmetric, Hamiltonians,
which always possess real eigenbases. We stress that the complex conjugation
symmetry can be relaxed by adding a detuning $\delta_{j}=\nu\pm\epsilon$
to one or more of the control lasers, which can be helpful if we aim
to quantum simulate a non-real Hamiltonian $H^{T}$.

\section{Controllability\label{sec:Controllability}}

We now discuss the controllability problem and show that, within the
constraints set in the symmetry section, the selected controls are
theoretically able to prepare any quantum state, in the limit of infinite
available resources. First of all, we argue that starting from the
state $\left|\psi_{\text{in}}\right\rangle \otimes|q\rangle_{a}$
it is possible to shift the system qubits state to any magnetization
sector. This is shown recursively, by highlighting that the state
$\left|\downarrow\right\rangle_{j}|q\rangle_{a}$ under the action of $H_{j}$
for a time equal to $t=\pi/(\tilde{\Omega}\sqrt{q+1})$ is completely
mapped into the state $\left|\uparrow\right\rangle_{j}|q+1\rangle_{a}$, ultimately
increasing the magnetization by +1. Secondly, we show that our controls
$H_{j}$ provide \textit{strong controllability} within each magnetization
sector $m$ (of dimension $d_{m}$) of real states, meaning that the
Lie algebra generated by $\{H_{j}\}$ contains the Lie algebra of
imaginary, hermitian matrices onto the $m$ sector, generators of
SO$(d_{m})$, as we will show later in this section. %in the Supplementary Material (SM).
From this observation, it follows that any two real states $|\psi_{1}\rangle|q\rangle_{a}$
and $|\psi_{2}\rangle|q\rangle_{a}$ can be dynamically connected,
i.e.,~$|\psi_{2}\rangle=e^{-iH't/\hbar}|\psi_{1}\rangle$ for some
$t$, simply because the effective Hamiltonian $H'=i|\psi_{1}\rangle\langle\psi'|-i|\psi'\rangle\langle\psi_{1}|$,
with $|\psi'\rangle\propto|\psi_{2}\rangle-|\psi_{1}\rangle\langle\psi_{1}|\psi_{2}\rangle$,
can be generated with our resources.

We now prove that, by means of linear combination and commutation,
we can generate the whole algebra of imaginary Hermitian matrices
for the sector $m$ by starting from the bare resources $H_{j}=i(\sigma_{j}^{+}a^{\dagger}-\sigma_{j}^{-}a)$.
We start by considering the commutator $i[H_{j},H_{j'}]=i(\sigma_{j}^{-}\sigma_{j'}^{+}-\sigma_{j}^{+}\sigma_{j'}^{-})$,
for sites $j\neq j'$, which is an endomorphism of magnetization sector
$m$. Moreover, it is easy to show that $i[H_{j},i[H_{j},H_{j'}]]=\sigma_{j}^{z}H_{j'}$,
for $j\neq j'$. By recursion and locality, it follows that we can
generate any string operator of the form 
\begin{equation}
%\tilde{H}=
i(\sigma_{j}^{-}\sigma_{j'}^{+}-\sigma_{j}^{+}\sigma_{j'}^{-})\prod_{k\neq j,j'}(\sigma^{z})^{q_{k}}
\end{equation}
for any binary string $q_{k}$. We run one additional commutator,
namely $i[(\sigma_{j}^{-}\sigma_{j'}^{+}-\sigma_{j}^{+}\sigma_{j'}^{-})\sigma_{k}^{z},(\sigma_{k}^{-}\sigma_{k'}^{+}-\sigma_{k}^{+}\sigma_{k'}^{-})]=2i(\sigma_{j}^{-}\sigma_{j'}^{+}-\sigma_{j}^{+}\sigma_{j'}^{-})(\sigma_{k}^{-}\sigma_{k'}^{+}+\sigma_{k}^{+}\sigma_{k'}^{-})$
where all sites are different. By recursion, we conclude that we can
generate operators $\bar{H}$ which are tensor product of one operator
$i(\sigma_{j}^{-}\sigma_{j'}^{+}-\sigma_{j}^{+}\sigma_{j'}^{-})$
at a pair of sites $j\neq j'$, any amount of $\sigma_{\ell}^{z}$,
and any amount of $(\sigma_{k}^{-}\sigma_{k'}^{+}+\sigma_{k}^{+}\sigma_{k'}^{-})$
(each one acting on separate sites). We remark that these operators
form a basis for the imaginary Hermitian endomorphisms of the symmetry
sector $m$. To see this, consider the canonical basis $i(|q_{1}q_{2}q_{3}\ldots\rangle\langle r_{1}r_{2}r_{3}\ldots|-H.c.)$,
where $\vec{q}$ and $\vec{r}$ are any two-bit strings with total
magnetization $m$. This operator can be easily decomposed in operators
of the form $\bar{H}$: for bits $j$ that do not flip ($q_{j}=r_{j}=\pm1$)
just insert a $(1\pm\sigma_{j}^{z})$ in the tensor product string,
while for pairs of bits that flip-flop, insert a $(\sigma_{k}^{-}\sigma_{k'}^{+}+\sigma_{k}^{+}\sigma_{k'}^{-})$,
except one, which is substituted by $i(\sigma_{j}^{-}\sigma_{j'}^{+}-\sigma_{j}^{+}\sigma_{j'}^{-})$.
This concludes the proof.

\section{Parameters optimization\label{sec:Parameters-optimization}}

In the section, we describe the method used for the optimization of
variational parameters $\boldsymbol{\theta}$ in the numerical simulations
of VQE with the circuits given in the main text. Since the purpose
of the paper is to study the achievable accuracy of the proposed approaches,
we employ the exact value of cost function $\langle H_{\text{SSH}}\rangle_{\boldsymbol{\theta}}$
in the optimization. The expectation value $\langle H_{\text{SSH}}\rangle_{\boldsymbol{\theta}}$
is obtained by simulating the circuits in terms of density matrixes
without considering shot noise caused by the finite number of measurements
of the trial ions states. To optimize parameters $\boldsymbol{\theta}$ we
use a gradient-based optimization algorithm, which relies on numerically
accessible approximate values of $\nabla_{\theta_{i}}\langle H_{\text{SSH}}\rangle_{\boldsymbol{\theta}}=(\langle H_{\text{SSH}}\rangle_{\boldsymbol{\theta}}\mid_{\theta_{i}\to\theta_{i}+\Delta}-\langle H_{\text{SSH}}\rangle_{\boldsymbol{\theta}})/\Delta$$\mid_{\Delta\to0}$.
To be more precise, we use the basin-hopping method~\cite{basinhopping,basinhoppingA},
which is a two-phase method that combines a global stepping algorithm
with local minimization at each step. As a local minimization algorithm,
we use the quasi-newton method of Broyden, Fletcher, Goldfarb, and
Shanno (BFGS)~\cite{BFGS1,BFGS2,BFGS3,BFGS4}.

To decrease noise in the data presented in the result section, we ensured
that $\langle H_{\text{SSH}}\rangle_{\boldsymbol{\theta}}$ is monotonically
decreases with decrease of temperature $T$ of the COM mode (or equivalently
number of excitations $n_{0}$), with decrease of state size $N_{\text{ions}}$,
and with growth of the number of variational parameters $N_{p}$.
We begin by independently optimizing parameters for all data points
(defined by one or two values of $n_{0}$, $N_{\text{ions}}$, and
$N_{p}$), each of which starts from several sets of initial parameters
$\boldsymbol{\theta}$ with values $\sim0.001-0.1$. Then, the optimized
parameters of the best result in one data point is used as the initial
parameters set for the neighbor data point, and the obtained result
is compared with the existing ones. In the case when the neighbor
data point requires less initial parameters, the extra parameters
are removed from the inserted parameters set; otherwise, parameters
with $0$ can be inserted. To obtain the presented data, we iteratively
re-optimize the parameters from one side of data points range to another
and, afterward, in the reverse direction until convergence.

\section{Gates sequence in the circuit\label{sec:Gates-sequence-in}}

\begin{figure}
\includegraphics[width=\linewidth]{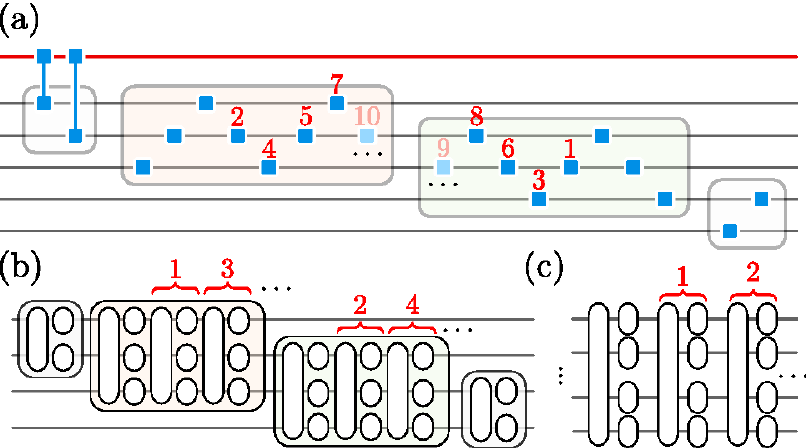}

\caption{Strategy for increasing of the number of the variational parameters
in circuits: (a) QDB-MPS, (b) CSD-MPS, and (c) CSA. The numbers indicate
the order in which the new variational operations are inserted. The
operations without the numbers represent the minimum set. In (a),
blue links in the left box represent interactions between QDB (red
line) and qubits (black lines) and, for simplicity, are replaced by
the blue squares in the following boxes.}
\label{fig:}
\end{figure}

In the section, we consider in detail the sequences of variational
operations, Fig.~\eqref{fig:}, used to construct circuits considered
in our work. The number of variational operations in the edge boxes
of the QDB-MPS and CSD-MPS circuits is fixed to $2$ and 3 correspondingly,
as shown in Fig.~\eqref{fig:}. In these circuits, the number of
variational parameters is increased by progressively inserting operations
in the first bulk box (orange) or second (green), alternately. In
the QDB-MPS circuit, we increase this number by 1, arranging the operations
such that the last operation in one box and the first in the next
box do not act to the same qubit. In the CSD-MPS circuit, we add a
single layer, consisting of a local MS operation and $\sigma_{i}^{z}$
rotations at each ion. To increase the number of variational operations
in the CSA circuit, we add a single layer consisting of $H_{XY}$
operations and $\sigma_{i}^{z}$ per ion.

The trial states in the CSA circuit are obtained as $\left|\psi_{\text{out}}(\boldsymbol{\theta})\right\rangle =\prod_{i}\text{exp}\{-i\theta_{i}H_{i}^{(r)}\}\left|\psi_{\text{in}}\right\rangle $,
with $\{H_{i}^{(r)}\}$ the resource operations. Afterward, the required
correlation functions can be obtained from $\left|\psi_{\text{out}}(\boldsymbol{\theta})\right\rangle $.
Since this method requires numerical operation with the full quantum
state, the number of the qubits in the CSA circuit in our study is
limited by $12$. On the contrary, one can see from Fig.~\ref{fig:}
that, in order to obtain correlation functions of the states generated by the
QDB-MPS and CSD-MPS, the full state is not required to be kept, but
only a reduced density matrix of at maximum $l$ qubits plus the QDB,
with $l$ the size of the bulk boxes. The qubits which do not more
participate in the evolution until the end of the circuit can be measured
in the required bases, and the next qubit can be added to the remained
qubits state by using the Kronecker product operation.

\section{Energy bounds on fidelity and purity\label{sec:Energy-bounds-on}}

In this section, we show that the final energy $\langle H\rangle=\text{Tr}[H\rho(\boldsymbol{\theta})]$,
in respect to a target Hamiltonian $H$, measured on the optimized
variational quantum state $\rho(\boldsymbol{\theta})$ imposes a bound on
its fidelity $F=\langle\psi_{\text{targ}}|\rho(\boldsymbol{\theta})|\psi_{\text{targ}}\rangle$,
with $|\psi_{\text{targ}}\rangle$ the exact ground state. Establishing
this bound requires a good estimate of the two lowest energy levels
$E_{0}$ and $E_{1}$, which can be estimated on the quantum device
via a subspace expansion technique for mixed states, illustrated in
detail in the Supplementary Information of Ref.~\cite{Colless2018}.
This analysis provides self-verification of the variational quantum
state preparation algorithm~\cite{Kokail2019}. Alternatively, for
1D lattice Hamiltonians, $E_{0}$ and $E_{1}$ can be obtained using numerical
method DMRG.

Let us consider the following inequality 
\begin{multline}
\langle H\rangle-E_{0}=\text{Tr}[H\rho(\boldsymbol{\theta})]-E_{0}\\
=\sum_{j\geq0}\langle e_{j}|(H-E_{0}I)\rho(\boldsymbol{\theta})|e_{j}\rangle=\sum_{j\geq1}(E_{j}-E_{0})\langle e_{j}|\rho(\boldsymbol{\theta})|e_{j}\rangle\geq\\
\geq(E_{1}-E_{0})\sum_{j\geq1}\langle e_{j}|\rho(\boldsymbol{\theta})|e_{j}\rangle\\
=(E_{1}-E_{0})\left(1-\langle e_{0}|\rho(\boldsymbol{\theta})|e_{0}\rangle\right)=(E_{1}-E_{0})\,(1-\mathcal{F})
\end{multline}
stating that the state fidelity $F$ is bound from below by the ratio
\begin{equation}
{F}\geq\frac{E_{1}-\langle H\rangle}{E_{1}-E_{0}},
\end{equation}
delivering an actual (positive) bound only when the variational state
energy is below the first excited energy, i.e. $\langle H\rangle<E_{1}$.
A lower bound on the fidelity, in turn, infers a bound on the purity
$\text{Tr}[\rho^{2}]$.
In fact 
\begin{multline}
\text{Tr}[\rho^{2}]=\sum_{j\geq0}\langle e_{j}|\rho^{2}|e_{j}\rangle\geq\langle e_{0}|\rho^{2}|e_{0}\rangle\\
=\sum_{j\geq0}\langle e_{0}|\rho|e_{j}\rangle\langle e_{j}|\rho|e_{0}\rangle\geq\langle e_{0}|\rho|e_{0}\rangle^{2}={F}^{2}.
\end{multline}
In conclusion, as long as $\langle H\rangle\le E_{1}$,
we have that 
\begin{equation}
\text{Tr}[\rho^{2}]\geq\left(\frac{E_{1}-\langle H\rangle}{E_{1}-E_{0}}\right)^{2},
\end{equation}
approaching $\text{Tr}[\rho^{2}]=1$ (certified pure state) in the
limit $\langle H\rangle\to E_{0}$.

\bibliographystyle{psi_quantum}
\bibliography{Bibliography}

\end{document}